\documentclass[5p,preprint,12pt]{elsarticle}

\makeatletter\if@twocolumn\PassOptionsToPackage{switch}{lineno}\else\fi\makeatother

\usepackage{tabulary,xcolor}
\usepackage{svg}
\usepackage{amsfonts,amsmath,amssymb}
\usepackage[T1]{fontenc}
\usepackage{gensymb}
\usepackage{siunitx}
\usepackage{dblfloatfix}
\usepackage[none]{hyphenat} 
\usepackage{ragged2e}
\usepackage{setspace} 
\usepackage[version=4]{mhchem}
\usepackage{geometry}
\usepackage[superscript]{cite}

\makeatletter
\let\save@ps@pprintTitle\ps@pprintTitle
\def\ps@pprintTitle{\save@ps@pprintTitle\gdef\@oddfoot{\footnotesize\itshape \null\hfill\today}}
\def\hlinewd#1{%
  \noalign{\ifnum0=`}\fi\hrule \@height #1%
  \futurelet\reserved@a\@xhline}

\AtBeginDocument{\ifNAT@numbers \biboptions{sort&compress}\fi}

\makeatother

\usepackage{ifluatex}
\ifluatex
\defaultfontfeatures{Ligatures=TeX}
\usepackage[]{unicode-math}
\unimathsetup{math-style=TeX}
\else 
\usepackage[utf8]{inputenc}
\fi 
\ifluatex\else\usepackage{stmaryrd}\fi

\usepackage{url,multirow,morefloats,floatflt,cancel,tfrupee}
\makeatletter

\AtBeginDocument{\@ifpackageloaded{textcomp}{}{\usepackage{textcomp}}}
\makeatother
\usepackage{todonotes}
\usepackage{colortbl}
\usepackage{xcolor}
\usepackage{pifont}
\usepackage[nointegrals]{wasysym}
\usepackage{atbegshi}
\makeatletter

\def\mcWidth#1{\csname TY@F#1\endcsname+\tabcolsep}

\def\cAlignHack{\rightskip\@flushglue\leftskip\@flushglue\parindent\z@\parfillskip\z@skip}
\def\rAlignHack{\rightskip\z@skip\leftskip\@flushglue \parindent\z@\parfillskip\z@skip}

\@ifundefined{etal}{}{}

\usepackage{ifxetex}
\ifxetex\else\if@twocolumn\@ifpackageloaded{stfloats}{}{\usepackage{dblfloatfix}}\fi\fi

\AtBeginDocument{
\expandafter\ifx\csname eqalign\endcsname\relax
\def\eqalign#1{\null\vcenter{\def\\{\cr}\openup\jot\m@th
  \ialign{\strut$\displaystyle{##}$\hfil&$\displaystyle{{}##}$\hfil
      \crcr#1\crcr}}\,}
\fi
}

\AtBeginDocument{%
  \@ifpackageloaded{endfloat}%
   {\renewcommand\efloat@iwrite[1]{\immediate\expandafter\protected@write\csname efloat@post#1\endcsname{}}}{\newif\ifefloat@tables}%
}%

\def\BreakURLText#1{\@tfor\brk@tempa:=#1\do{\brk@tempa\hskip0pt}}
\let\lt=<
\let\gt=>
\def\processVert{\ifmmode|\else\textbar\fi}

\@ifundefined{subparagraph}{
\def\subparagraph{\@startsection{paragraph}{5}{2\parindent}{0ex plus 0.1ex minus 0.1ex}%
{0ex}{\normalfont\small\itshape}}%
}{}

\newcommand\role[1]{\unskip}
\newcommand\aucollab[1]{\unskip}
  
\@ifundefined{tsGraphicsScaleX}{\gdef\tsGraphicsScaleX{1}}{}
\@ifundefined{tsGraphicsScaleY}{\gdef\tsGraphicsScaleY{.9}}{}
\def\checkGraphicsWidth{\ifdim\Gin@nat@width>\linewidth
	\tsGraphicsScaleX\linewidth\else\Gin@nat@width\fi}

\def\checkGraphicsHeight{\ifdim\Gin@nat@height>.9\textheight
	\tsGraphicsScaleY\textheight\else\Gin@nat@height\fi}

\def\fixFloatSize#1{}
\let\ts@includegraphics\includegraphics

\def\inlinegraphic[#1]#2{{\edef\@tempa{#1}\edef\baseline@shift{\ifx\@tempa\@empty0\else#1\fi}\edef\tempZ{\the\numexpr(\numexpr(\baseline@shift*\f@size/100))}\protect\raisebox{\tempZ pt}{\ts@includegraphics{#2}}}}

\AtBeginDocument{\def\includegraphics{\@ifnextchar[{\ts@includegraphics}{\ts@includegraphics[width=\checkGraphicsWidth,height=\checkGraphicsHeight,keepaspectratio]}}}

\DeclareMathAlphabet{\mathpzc}{OT1}{pzc}{m}{it}

\def\URL#1#2{\@ifundefined{href}{#2}{\href{#1}{#2}}}

\def\UrlOrds{\do\*\do\-\do\~\do\'\do\"\do\-}%
\g@addto@macro{\UrlBreaks}{\UrlOrds}

\edef\fntEncoding{\f@encoding}

\makeatother

\newif\ifmultipleabstract\multipleabstractfalse%
    \makeatletter
\def\ead{\@ifnextchar[{\@uad}{\@ead}}
\gdef\@ead#1{\bgroup
   \def\_{\string\underscorechar\space}
   \def\{{\string\lbracechar\space}
   \def\textdagger{\string\textdagger\space}
   \def\texttildeapprox{\string\texttildeapprox\space}
   \def~{\hashchar\space}
   \def\}{\string\rbracechar\space}
   \edef\tmp{\the\@eadauthor}
   \immediate\write\@auxout{\string\emailauthor
     {#1}{\expandafter\strip@prefix\meaning\tmp}}
  \egroup
}
\gdef\emailauthor#1#2{\stepcounter{ead}
      \g@addto@macro\@elseads{\justifying
      \let\corref\@gobble
      \eadsep\texttt{#1} (#2)
      \def\eadsep{\unskip,\space}}
}

\makeatother
\usepackage[scaled]{helvet}
\usepackage[T1]{fontenc}

\begin{document}
\begin{frontmatter}

\title{Correlative Microstructural Analysis of a Weathered Nantan Meteorite Fragment}
\author[1]{Graeme J. Francolini\corref{correspondingauthor}}
\cortext[correspondingauthor]{Corresponding author.}
\ead{graeme.francolini@ubc.ca}
\author[2]{Brendan V. Dyck}
\author[3]{Paul Mack}
\author[1]{Ben Britton}
\ead{ben.britton@ubc.ca}
\affiliation[1]{Department of Materials Engineering, University of British Columbia, BC, Canada}
\affiliation[2]{Department of Earth and Environmental Sciences, University of British Columbia, Kelowna, BC, Canada}
\affiliation[3]{Thermofisher Scientific}
\date{\today}                     

\begin{abstract}

\par The weathering of iron-rich phases within meteorites is a multi-stage process that significantly alters the microstructure and chemical composition based both on the environmental condition at the location of landing and the exposure time since the fall. This work investigates the resulting phases of this process in a correlative and comparative manner using a naturally weathered Nantan meteorite fragment. Techniques including X-ray Photoelectron Spectroscopy (XPS), Energy Dispersive X-ray Spectroscopy (EDS), and X-ray Fluorescence Spectroscopy (XRF) were used for chemical and compositional determination and X-ray Diffraction (XRD) and Electron Backscatter Diffraction (EBSD) for phase determination and microstructural analysis. 

\par The combination of these techniques revealed the meteorite matrix to be predominantly composed of magnetite, with goethite and maghemite present as secondary phases. The meteorite had two pairs of distinct regions, one separated by Ni content and the other by grain size and Cl distribution. The Ni content regions are believed to have formed from a former Widmanstätten pattern, with the high ($\geq$\SI{2.6}{at\percent}) and low ($\leq$\SI{0.9}{at\percent}) Ni regions from plessite and kamacite, respectively. The kamacite and taenite within the plessite are suggested to have weathered through distinct mechanisms. The grain size regions were separated by a visible boundary 100-\SI{200}{$\mu$m} wide where grains transitioned rapidly from fine to elongated grains and the Cl content changed from low and dispersed to concentrated spots.

\par Lastly, a brecciated phase was found within the sample and was identified as a large cohenite grain which exhibited signs of weathering, including vein-like extensions of the matrix composed of NiO and magnetite, and deposits of iron and nickel carbonates. 

\end{abstract}

\end{frontmatter}
\AtBeginShipoutNext{\AtBeginShipoutDiscard}
\onecolumn   
\pagenumbering{arabic}
\section{Introduction}
\par To understand the fall-to-find timeline of meteorites and how best to store and preserve these rare extra-terrestrial materials it is important to understand the weathering of meteorites, including how it occurs and how it changes the material. This is especially relevant for iron meteorites including the IAB complex iron meteorites. This complex of meteorites are believed to have formed from molten pools rather than the molten cores of their parent asteroid and consist primarily of iron-nickel alloys species (kamacite, taenite, tetrataenite) and silicate inclusions such as olivine and plagioclase \cite{Scott2020}. However, these meteorite phases may undergo chemical modification through weathering, both terrestrially and in space. Often weathering results in a notable change in the metallic and mineral phases present, as well as changes in trace elemental compositions and microstructure of meteorites. 

\par The primary objective of this work is to investigate the chemical and microstructural modification which occurs in iron meteorites from this weathering process, through a multi-modal microstructural and chemical analysis using Energy Dispersive X-ray Spectroscopy (EDS), X-ray Diffraction (XRD), X-ray Photoelectron Spectroscopy (XPS), X-ray Fluorescence spectroscopy (XRF) and Electron Backscatter Diffraction (EBSD), with EBSD providing microstructural information and assisting in phase identification. These techniques are used to explore the changes in the elemental composition, phases and microstructure present in a weathered Nantan iron meteorite and to relate these to the weathering mechanism which formed them.

\par According to the literature, the terrestrial weathering process of iron-based phases is proposed to occur through two possible mechanisms. The first mechanism is a multi-step process, with Fe-Ni phases transitioning to the chlorine stabilized oxyhydroxide, akaganéite, followed by decomposition into other iron oxide and oxyhydroxide species (Eq. \ref{Eqn:AkaganeiteDecomA} and \ref{Eqn:AkaganeiteDecomB}) \cite{Binzel2006}.
\begin{equation}
    \label{Eqn:AkaganeiteDecomA}
    \ce{{\underbrace{30 Fe^0 + 2Ni^0}_{kamacite}} + 47O + 19H_2O + 4H^+_{(aq)} + 4Cl^-_{(aq)} ->[irreversible] 2{\underbrace{[Fe_{15}Ni][O_{12}(OH)_{20}]Cl_2(OH)}_{akaganéite}}} 
\end{equation}
\begin{equation}
    \label{Eqn:AkaganeiteDecomB}
        \ce{2{\underbrace{[Fe_{15}Ni][O_{12}(OH)_{20}]Cl_2(OH)}_{akaganéite}} ->
        {\underbrace{7\gamma-Fe_2O_3}_{maghemite}} + {\underbrace{16\alpha-FeOOH}_{goethite}} +2NiO + 11H_2O + 4H^{+}_{(aq)} + 4Cl^-_{(aq)}}
\end{equation}
The second mechanism is the conversion of the Fe-Ni metal into magnetite through dissolution and subsequent hydrolysis (Eq. \ref{Eqn:MagnetiteFormationA}) and oxidation (Eq. \ref{Eqn:MagnetiteFormationB}) of dissolved Fe\textsuperscript{2+} and Fe\textsuperscript{3+} ions, respectively \cite{Bland1998a, Buchwald1989, Faust1973}. 
\begin{equation}
    \label{Eqn:MagnetiteFormationA}
        \ce{Fe^{2+}_{(aq)} +2H2O -> Fe(OH)_2 + 2H+_{(aq)}}
\end{equation}
\begin{equation}
    \label{Eqn:MagnetiteFormationB}
        3Fe(OH)_2 \ce{->} \underbrace{Fe_3O_4}_{magnetite} + H_2 + 2H_2O
\end{equation}
\par Furthermore, the extent of weathering experienced by meteorites is affected by the climate of the fall site. Previous work by Bland \emph{et al.} \cite{Bland1998a} has shown a relation between the presence of magnetically ordered species (magnetite and maghemite) and paramagnetic species (akaganéite, goethite, lepidocrocite) with the climate which chondrite meteorites landed. A high abundance of magnetically ordered species were found in meteorites which landed in humid regions; whereas magnetite is uncommon for Antarctic samples \cite{Bland1998a}. Additionally, ``hot'' desert meteorites have been found to contain higher contents of sulfates, Ca-carbonate, and silica as compared to Antarctic finds \cite{Binzel2006, Lee2004}. The difference in weathering climate also results in differences within the post-weathering microstructure. Both hot and cold climates result in predominantly Fe,Ni metal-based modification which forms vein-like structures in the metal surface. However, weathering of ``hot'' desert meteorites causes greater amounts of brecciation and dissolution compared to cold desert weathering \cite{Lee2004}.      

\par This brecciation and vein-like microstructure has been observed previously through the use of scanning electron microscopy (SEM) in both chondrite and iron meteorites \cite{VanGinneken2022, Gurdziel2015}. These vein-like structures are typically observed intersecting or surrounding (Fe,Ni) metal surfaces. These are composed of iron oxides and oxyhydroxides, due to dissolution of high-Fe content phases and subsequent precipitation of dissolved Fe. Individual iron oxyhydroxide species have also been observed under SEM as distinct phases that formed during low-temperature recrystallization. Goethite may appear as a globular \cite{VanGinneken2022} or large and angular platelet \cite{Karwowski2009} structure. Akaganéite is found to have a ``cigar-shaped'' or ``cotton-ball'' structure \cite{VanGinneken2022}. Lastly, lepidocrocite is forms as thin and fine plates \cite{VanGinneken2022, Karwowski2009}. 

\par Prior work has identified these weathering phases have been identified using EDS, Mössbauer spectroscopy \cite{VanGinneken2022, BenderKoch1994} and electron probe microanalysis (EPMA) \cite{Lee2004}. However, in these works the elemental composition of Fe, Ni, and O can vary significantly between individual samples and weathered metallic regions in a sample. In the work of Van Ginneken \emph{et al.} samples that underwent artificial weathering had variation of 22.9 - \SI{44.4}{wt\%O}, 49.2 - \SI{67.5}{wt\%Fe}, and 1.6 - \SI{15.0}{wt\%Fe} \cite{VanGinneken2022}. For oxides, this variation has been seen to be 58.4\text{-}\SI{84.1}{wt\%Fe\textsubscript{2}O\textsubscript{3}} and 0.6 - \SI{15.3}{wt\%NiO} \cite{Lee2004}. 

\par Significant work has also been performed to measure the composition of meteorites and to determine how their composition is altered through weathering. A variety of X-ray based methods are used to determine the elemental composition, such as EDS, XPS, and XRF. These techniques are rarely used alone, instead they are often paired with other techniques to provide a better understanding of a specific material property. For example, EDS is typically used as a complementary technique alongside SEM and XRD to identify phases or evaluate the composition of elementally similar but polymorphically distinct phases \cite{Goryunov2023}. EDS has also been used with XRF and X-ray microtomography to provide a three dimensional structural, compositional, and phase understanding of meteorites \cite{Machado2025}. Additionally, XRF has the unique benefit that can be used as a handheld characterization tool, allowing for bulk analysis with minimal or no sample preparation \cite{Allegretta2020, Gemelli2015}, and potentially deployed in the field.   

\par This work looks to showcase the benefits of multi-modal characterization through the combination of EDS, XPS, EBSD, XRD, and XRF for complex materials like meteorites. Secondly, to use these techniques to understand the distinct Ni and subsurface-Cl compositional regions which formed during the weathering of this Nantan meteorite fragment and how the resulting microstructure has been influenced by the formation of these regions. 


\section{Materials and Methods}

\subsection{Material Preparation}

\par
\begin{figure}[!hbt]
    \centering
    \includegraphics[width=0.9\linewidth]{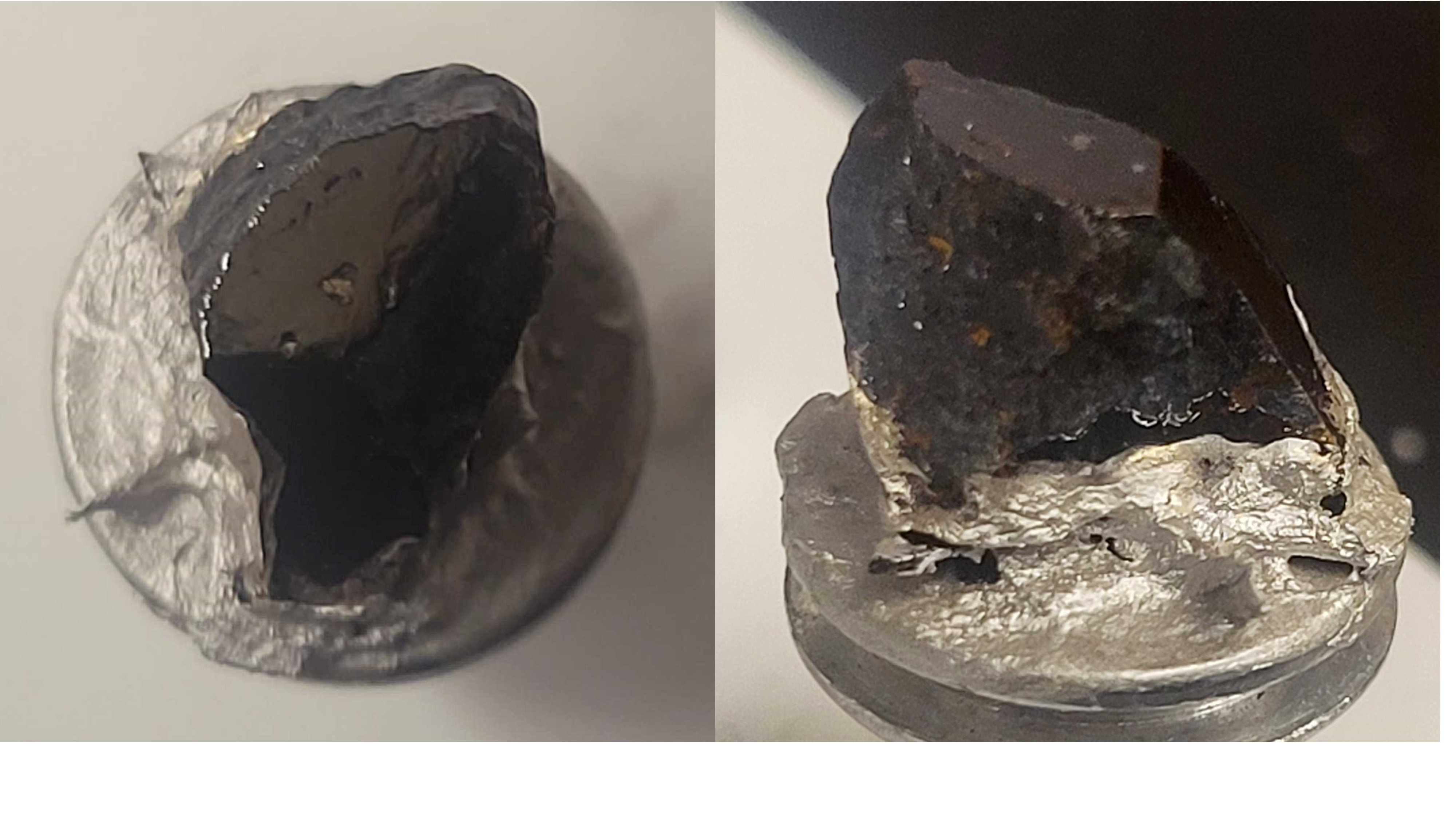}
    \caption{Naturally weathered Nantan meteorite fragment secured to a standard 
    \SI{12.5}{mm} SEM stub, using conductive silver paint and super glue, with the surface of the sample and stub co-parallel. Images show the analysis surface (left) and the weather exterior of the sample (right).}
    \label{Fig:SampleImage}
\end{figure}

 A Nantan meteorite sample (Figure \ref{Fig:SampleImage}) was sourced from Rocks $\&$ Gems Canada, Whistler, BC, Canada. This sample had a dull dark colour and a bulbous surface, with small spots of orange rusting. The sample was secured to a standard SEM stub using silver paint and super glue, then metallographically polished using an Allied High Tech Multiprep Precision Polisher, with active sample rotation and oscillation. The sample was prepared using 600, 800, and 1200 grit SiC, followed by 6 and \SI{1}{$\mu$m} monocrystalline diamond suspensions and finally a \SI{0.05}{$\mu$m} colloidal alumina/silica suspension. Each polishing step was performed at \SI{250}{RPM} for 15 minutes, with 3 minute on pad water cleaning steps performed between the \SI{6}{$\mu$m} - \SI{0.05}{$\mu$m} steps. Finally, the sample was rinsed with ethanol and dried using hot air.

\subsection{Material Analysis}

\par A variety of analysis techniques and systems were used to perform microstructural and chemical analysis of the Nantan sample. For simplicity, this section has been separated into the individual systems used for the analysis. 

\subsubsection{Electron-based Elemental Mapping and Imaging}

\par
\begin{figure}[!hbt]
    \centering
    \includegraphics[width=0.9\linewidth]{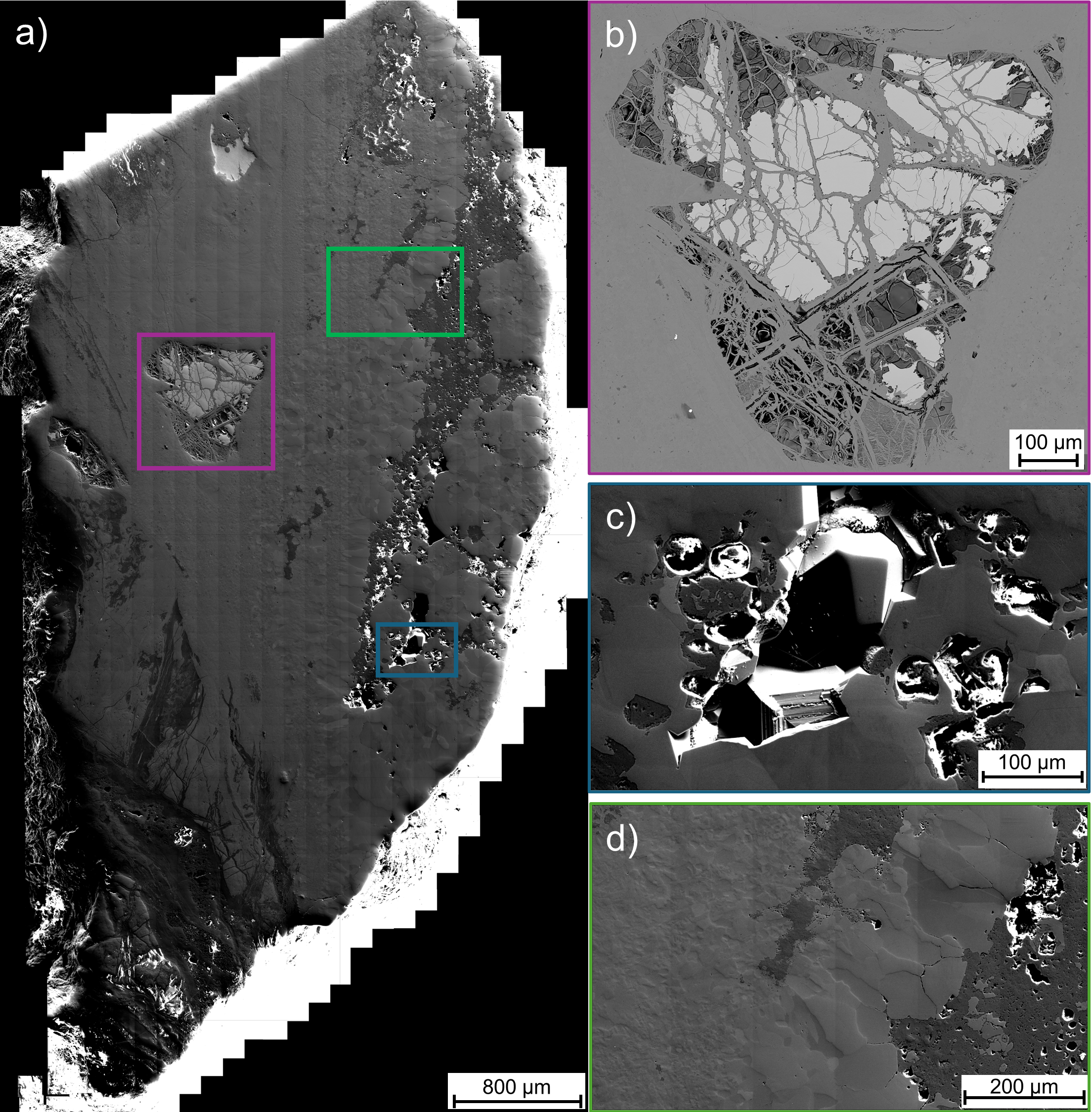}
    \caption{a) Large area secondary electron map of polished Nantan meteorite sample. b) Backscatter electron image of large inclusion present on meteorite surface (purple box). c) Large faceted crystals present within cracks on the sample surface (blue box). d) Grain structure transition region from small grains to large elongated grains. C) and d) are subset images taken from the larger stitched map (a).}
    \label{Fig:LargeStitchedMap}
\end{figure}

 Secondary electron (SE), backscatter electron (BSE) microscopy, and EDS mapping were performed on an Apreo 2 ChemiSEM (ThermoFisher Scientific). Data was collected at \SI{20}{kV} with a probe current of \SI{4}{nA}. The MAPS software allows for an integrated workflow and large area ``montage''-based image stitching. Stitching was performed in MAPS with a \SI{10}{\percent} tile overlap. The following maps were collected:

\begin{enumerate}
    \item Large area (Fig. \ref{Fig:LargeStitchedMap} a) - 27x60 tile montage, \SI{0.293}{$\mu$m} pixel size and \SI{26.51}{$\mu$s} dwell time. EDS mapping, BSE, and SE, data was collected.
    \item Higher resolution (Fig. \ref{Fig:LargeStitchedMap} b) - 18x20 tile montage, \SI{0.1}{$\mu$m} pixel size and \SI{88.4}{$\mu$s} dwell time.
\end{enumerate}

\par The EDS data was collected with a Truesight Pro 70 detector (ThermoFisher Scientific) and quantified in MAPS. SE was collected with the Everhart-–Thornley (ET) detector. BSE was collected with the in-column trinity T1 detector.   

\subsubsection{Local Phase Identification + Crystallographic Mapping}

\par Further electron microscopy was performed on a Tescan AMBER-X plasma focused ion beam-scanning electron microscope (pFIB-SEM). All data was captured using a \SI{10}{nA} beam current and  \SI{20}{kV} accelerating voltage. EBSD was performed using an Oxford Instruments Symmetry S2 detector and the sample loaded into a \SI{70}{\degree} pre-tilted sample holder. EDS data was captured using an Oxford Instruments Ultim Max 170 detector and the Pulse Pile up correction was enabled. 

EBSD was captured both independently and in tandem with EDS using Oxford Instruments PhaseID method. The independent EBSD data was captured at a working distance of \SI{13.8}{mm}, an acquisition speed of \SI{1393.7}{Hz}, a resolution of 156x\SI{128}{pixels}, a step size of \SI{0.5}{$\mu$m}, and an exposure time of \SI{0.7}{ms}. 

Three regions of interest were analyzed using a combination of EDS and EBSD to identify the possible material phases present in the meteorite surface. The specific parameters for each region are given in table \ref{Tbl:PhaseIDParams}. Post-acquisition analysis of the data was performed in Aztec. Additional analysis of the EBSD data was performed in Aztec Crystal using MapSweeper which performed a `template matching' approach to compare the captured diffraction patterns to a dynamical-diffraction based simulation of each identified phase, thus improving the indexing precision \cite{Winkelmann2020}. Followed by cross correlation coefficient filtering on an individual phase basis using the MTEX 5.10.2 package \cite{MTEX}.

\begin{table}[!hbt]
    \centering
    \begin{tabular}{|c|c|c|c|}
    \hline
        Region of Interest & 1 & 2 & 3 \\ \hline
        Working Distance (mm) & 6.8 & 13.4 & 6.4 \\ \hline
        Acquisition Speed (Hz) & 833.28 & 576.30 & 1136.89 \\ \hline
        Resolution (pixels) & 622x512 & 1244x1024 & 156x128 \\ \hline
        Step Size (\unit{$\mu$\meter}) & 5 & 1 & 0.5 \\ \hline
        Exposure Time (ms) & 3.0 & 1.5 & 7.1 \\ \hline
    \end{tabular}
    \caption{EBSD data capture parameters for phase identification.}
    \label{Tbl:PhaseIDParams}
\end{table} 

\subsubsection{X-ray Diffraction - Bruker D6 PHASER}

\par XRD was performed using a Bruker D6 Phaser Tabletop XRD with a LYNXEYE XE-T detector and a Cu $\kappa-\alpha$ anode. The polished face of the Nantan sample was analyzed in a coupled Bragg-Bretano ($\theta$--2$\theta$) scan at an anode voltage of \SI{40}{kV} and current of \SI{15}{mA}, with a continuous PSD fast scan mode. Data was collected over a $2\Theta$ range of \SI{10}{\degree} to \SI{140}{\degree}.

\subsubsection{X-ray Photoelectron Spectroscopy - Thermo Scientific Nexsa G2}

\par XPS spectra were acquired using a Thermo Scientific Nexsa G2 system with a Al K-Alpha X-ray source. The sample was cleaned using Ar\textsuperscript{+} sputtering at \SI{1}{keV}. Two areas of the sample were analyzed, the inclusion region of interest shown in \ref{Fig:LargeStitchedMap} b) and the sample matrix. Each of these areas were analyzed using a general XPS survey followed by higher resolution single element (Fe, O, Ni) scans. 

The survey of the matrix was performed in a fixed analyzer transmission mode and a constant analyzer energy (CAE) of 200. A total of five scans were performed with a \SI{1}{eV} pass energy and microfocus X-ray spot size of \SI{200}{$\mu$m}. The individual element scans were performed with a CAE of 50, \SI{0.1}{eV} pass energy and a step and spot size of \SI{200}{$\mu$m} over 10, 10, and 8 scans for Fe2p, O1s, and Ni2p respectively. All parameters for analysis of the inclusion region (overall and individual elements) were kept the same as for the matrix, except the number of scans for Fe2p which was increased to 9 scans.

Additional XPS data captured on a Kratos Analytical Axis Supra\textsuperscript{+} is provided in the Supplemental Information.

\subsubsection{X-ray Fluorescence Spectroscopy - IXRF Atlas $\mu$-XRF}

\par XRF analysis was performed using an IXRF Atlas $\mu$-XRF system with a Rh anode and four Amptek silicon drift detectors (each with \SI{70}{mm^2} active areas). A large area scan of the entire polished sample surface was performed with a current of \SI{800}{nA}, \SI{50}{kV} voltage, \SI{20}{$\mu$m} step size, \SI{20}{$\mu$m} spot size and a \SI{10}{ms} dwell time. Spectral data was then extracted from the overall scan data for three regions of interest (11x11 pixels) where the Fe, Mn and Ni content were observed to vary significantly.

\section{Results}

\par Preliminary investigation of the Nantan fragment using SEM showed a large variety in the phases and structural morphology of the sample surface. As seen in Fig. \ref{Fig:LargeStitchedMap}, the different phases of the material appear to be sequestered in discrete areas of the sample. The upper left region of Fig. \ref{Fig:LargeStitchedMap} a) shows a finer grained matrix with faint vein-like structure and large BSE-bright phases present (Fig. \ref{Fig:LargeStitchedMap} b). While the matrix on the right side of the sample was made up of larger grains of varied sizes as well as voids in the surface that are infilled with an assemblage of crystalline phases (Fig. \ref{Fig:LargeStitchedMap} c). The matrix grain structure was also observed to vary along the horizontal centre of the sample, with a distinct 100-\SI{200}{$\mu m$} transition band from smaller equiaxed grains on the left to large elongated grains on the right in Fig. \ref{Fig:LargeStitchedMap} d).   

\subsection{Matrix}

\par
\begin{figure}[!hbt]
    \centering
    \includegraphics[width=\linewidth]{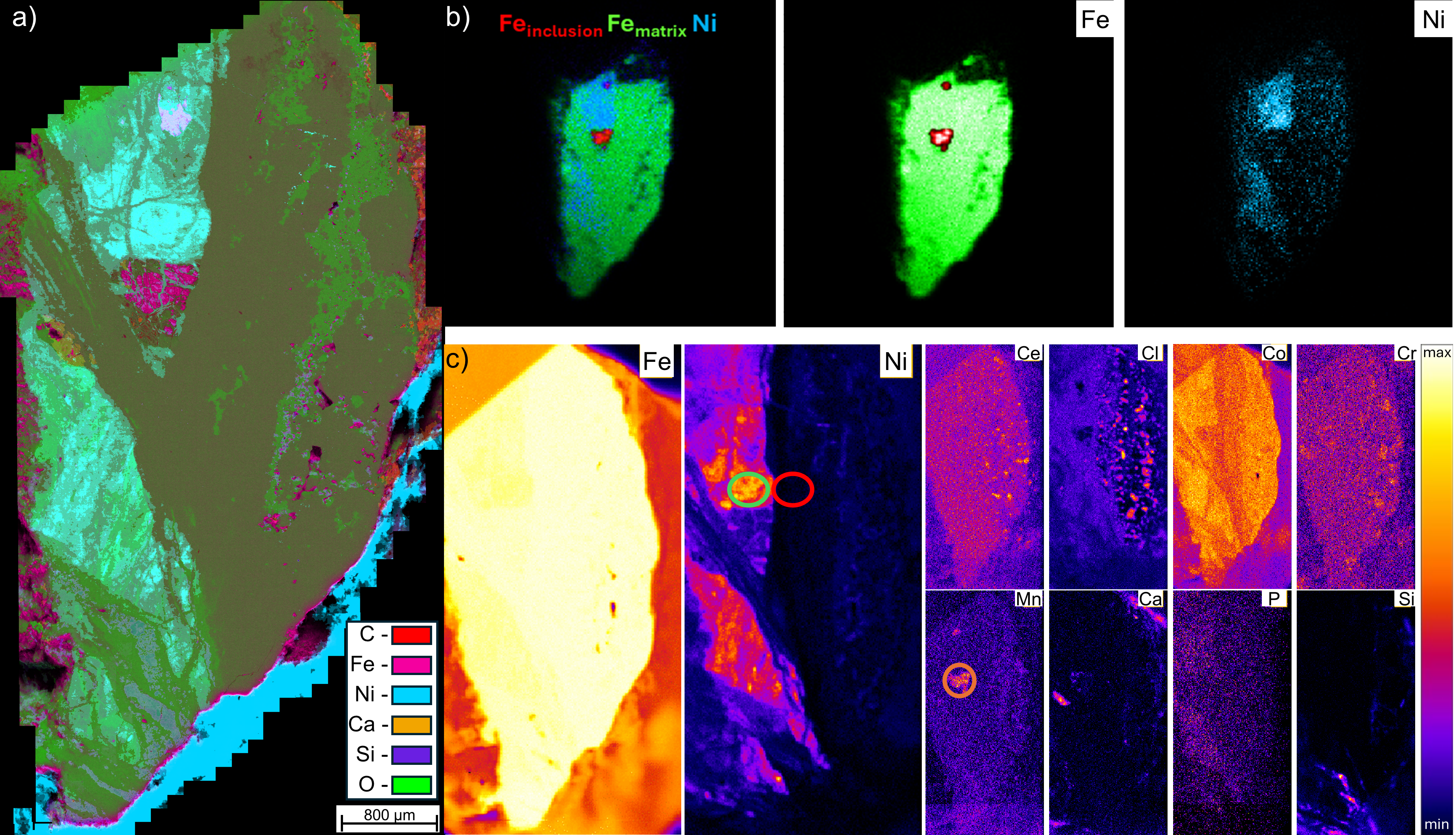}
    \caption{a) Large area EDS map of Nantan meteorite sample showing the distribution of detected elements. b) XPS SnapMaps of Nantan fragment showing the distribution of Fe and Ni. c) XRF maps of individual elements detected, with a focus on Fe and Ni. Three regions of interest are circled which were analyzed further, high Ni (green), low Ni (red), and the large inclusion (orange).}
    \label{Fig:LargeMapChemistry}
\end{figure}

 The EDS spectra (Fig. \ref{Fig:LargeMapChemistry} a) showed the sample surface was predominantly Fe, Ni, Ca, Si, and O. The sample consisted primarily of iron oxides of varying stoichiometry and regions of varying Ni content. Si was observed near the multiphase crystalline region shown in Fig. \ref{Fig:LargeStitchedMap} c) and concentrated in the BSE-darker region on the lower left of the sample. Ca was primarily observed as vein-fill within the brecciated grain in Fig. \ref{Fig:LargeStitchedMap} b). Additional analysis using XPS (Fig. \ref{Fig:LargeMapChemistry} b) and XRF (Fig. \ref{Fig:LargeMapChemistry} c) corroborated the presence of regions with high Ni concentration. It is believed that these regions of high and low Ni concentration are an artifact of a former Widmanstätten pattern microstructure present within the meteorite fragment before weathering. The low Ni regions forming from kamacite bands and the high Ni regions forming from former plessite (kamacite+taenite).

\par Compositional mapping of the sample surface using XRF revealed the presence of six other trace elements not seen using EDS. Those being, Ce, Cl, Co, Cr, Mn, and P. Co and P were seen to have higher concentrations within the high Ni region. Cl, Cr, and Ce were found to be generally distributed over the sample surface, with elevated concentrations in the low Ni region, with clustering right side of the sample surface. Mn was found to be concentrated in large inclusions present on the surface, like shown in Fig. \ref{Fig:LargeStitchedMap} b).

\par
\begin{figure}[!hbt]
    \centering
    \includegraphics[width=0.9\linewidth]{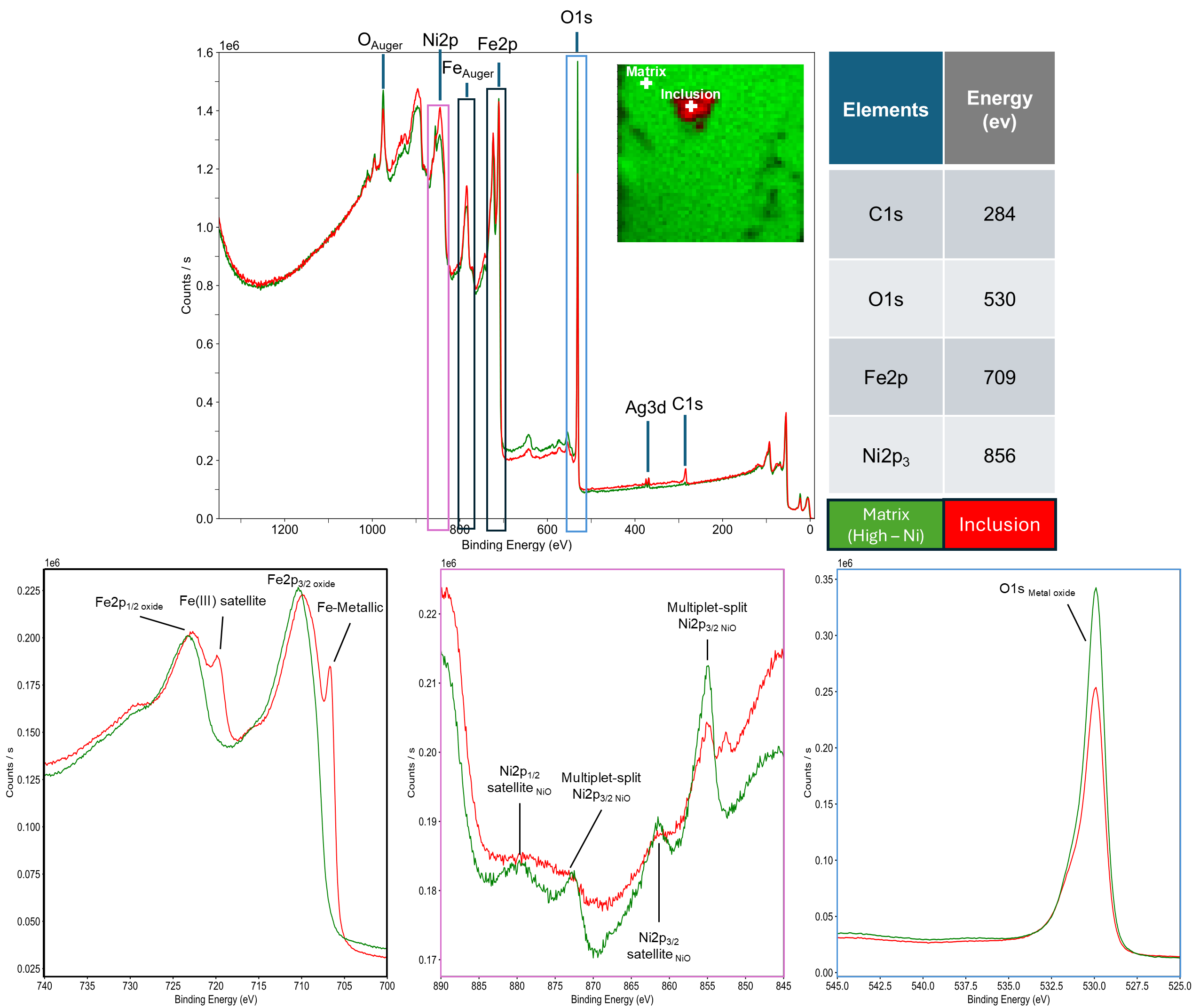}
    \caption{Overlayed XPS spectra taken at the large inclusion (red) and high Ni region of the matrix (green) with the major detected elements and their respective peak energies given in the included table (top right). Overlayed XPS spectra for XPS surveys of Fe, Ni, and O.}
    \label{Fig:XPSResults}
\end{figure}

 The XPS maps (Fig. \ref{Fig:LargeMapChemistry} b) showed the same spatial distribution of the Ni and Fe content as was found using EDS. However, XPS additionally suggested the presence of Fe\textsubscript{3}O\textsubscript{4} (magnetite), $\gamma$-Fe\textsubscript{2}O\textsubscript{3} (maghemite) and NiO (bunsenite) within in the matrix (Fig. \ref{Fig:XPSResults}). These additional XPS measurements were performed for the high Ni matrix region and the inclusion. The spectra of both regions showed peaks for Ni, Fe, O and Ag with the inclusion region having an additional peak for C. The detection of carbon within the inclusion and not the surrounding matrix matches the elemental mapping in Fig. \ref{Fig:LargeMapChemistry} a). Individual surveys for Fe, Ni, and O of the high Ni matrix region showed binding energies correlating to Fe\textsubscript{3}O\textsubscript{4} (magnetite) and NiO (bunsenite). The inclusion region showed a similar survey for Fe, though with additional Fe(III) satellite and Fe-metallic peaks. The Fe(III) satellite peak signifies the iron oxide is likely maghemite rather than magnetite, as the Fe(III) satellite peak is masked in magnetite \cite{Grosvenor2004}. The Fe-metallic peak is likely an artifact of the Ar\textsuperscript{+} sputter cleaning that was performed on the sample, as this cleaning can cause iron oxides to be reduced, forming metallic iron \cite{Grosvenor2004, Steinberger2014}. The trace amounts of Ag present is likely due to the silver paint used to secure the sample for analysis. Lastly, the peak seen at $\sim$\SI{784}{eV} is believed to be an Fe auger peak, rather than a Co peak which can appear between 778-\SI{780}{eV} \cite{Biesinger2011}. The surrounding peaks and their intensities have a better correlation to Fe than Co, as can be seen in the supplementary material provided in the work of Bagus \emph{et al.} \cite{Bagus2021}. 

\par
\begin{figure}[!thb]
    \centering
    \includegraphics[width=\linewidth]{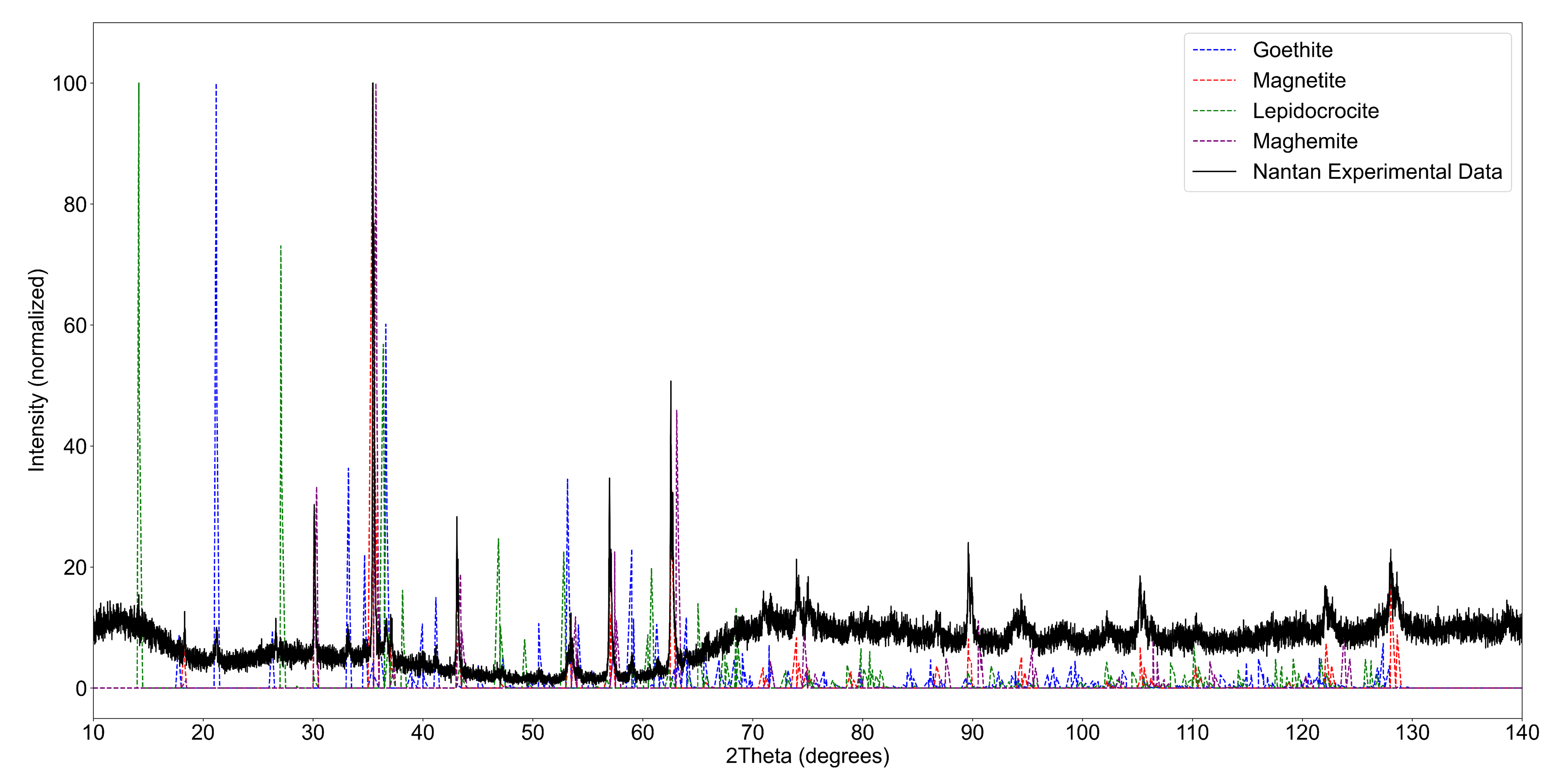}
    \caption{Experimentally obtained XRD diffractogram stacked on simulated XRD diffractograms from common minerals in iron meteorites. The CIF for each simulated pattern listed in the legend are provided as supplementary material.}
    \label{Fig:XRDSpectra}
\end{figure}

 The next step of characterizing the sample was to determine what mineral phases were present in the matrix. This was done in two stages: (1) using XRD to obtain a range of possible mineral phases; and (2) using a combination of EDS and EBSD through Oxford Instrument's PhaseID method. This method allows for the identification of unknown phases by comparing the elemental composition and electron backscatter pattern of a phase and comparing it to commercially available databases. The experimentally obtained XRD diffractogram is given in Fig. \ref{Fig:XRDSpectra} alongside XRD diffractograms simulated using VESTA \cite{Momma2011}. The simulated mineral diffractograms in Fig. \ref{Fig:XRDSpectra} were selected to be shown due to their common presence within iron meteorites and their correlation to experimental peaks. Comparisons with other minerals common to iron meteorites provided in the Supplementary Materials.

\par Comparison of the Nantan XRD diffractogram showed a strong correlation to magnetite and maghemite, though these are not easily discerned due to their similar XRD diffractograms. A weaker correlation was also seen for goethite ($\alpha$-FeO(OH)). It is possible that there is a correlation to lepidocrocite. However this cannot be confidently said due to the low intensity of the overlapped experimental peaks. This may be due to lower quantities of lepidocrocite and because the XRD analysis was performed on a singular polished face of the Nantan fragment, risking preferential phase identification. This was done as the sample would need to be powdered to alleviate this issue and the sample is required for further analysis. The Nantan fragment was also confirmed to be magnetic, further corroborating that the sample composition is a majority magnetite. 

\par The Fe composition of the matrix was found to vary between the high and low Ni regions, with the O and C concentrations measured with EDS staying relatively unchanged. The matrix surrounding the inclusion region was found to have a lower Ni concentration compared to the high Ni matrix region, but was consistently identified to have a higher Ni content than the low Ni matrix regions using all methods.

\par The elemental composition for the high Ni matrix region and the inclusion to be very similar for EDS and XPS. Though the Ni and Fe concentrations were found to be notably higher using XRF. This is anticipated as XRF was unable to identify the presence of O or C. Additionally, XRF has a significantly greater interaction (excitation) depth than electron-beam EDS, probing deeper into the bulk material rather than being limited to near-surface regions, which can further contribute to differences in the measured elemental concentrations. These elements have been confirmed to be present using the other analysis methods and therefore the lack of these elements in the quantification of the XRF skews the Ni and Fe composition to be higher than in reality. For example, the Fe concentration of the high Ni region is 91.7, 90.0, and 87.73 (at$\%$) when O and C are omitted, for XPS, EDS, and XRF respectively.   

\subsection{Surface Features}

\par
\begin{figure}[!thb]
    \centering
    \includegraphics[width=0.8\linewidth]{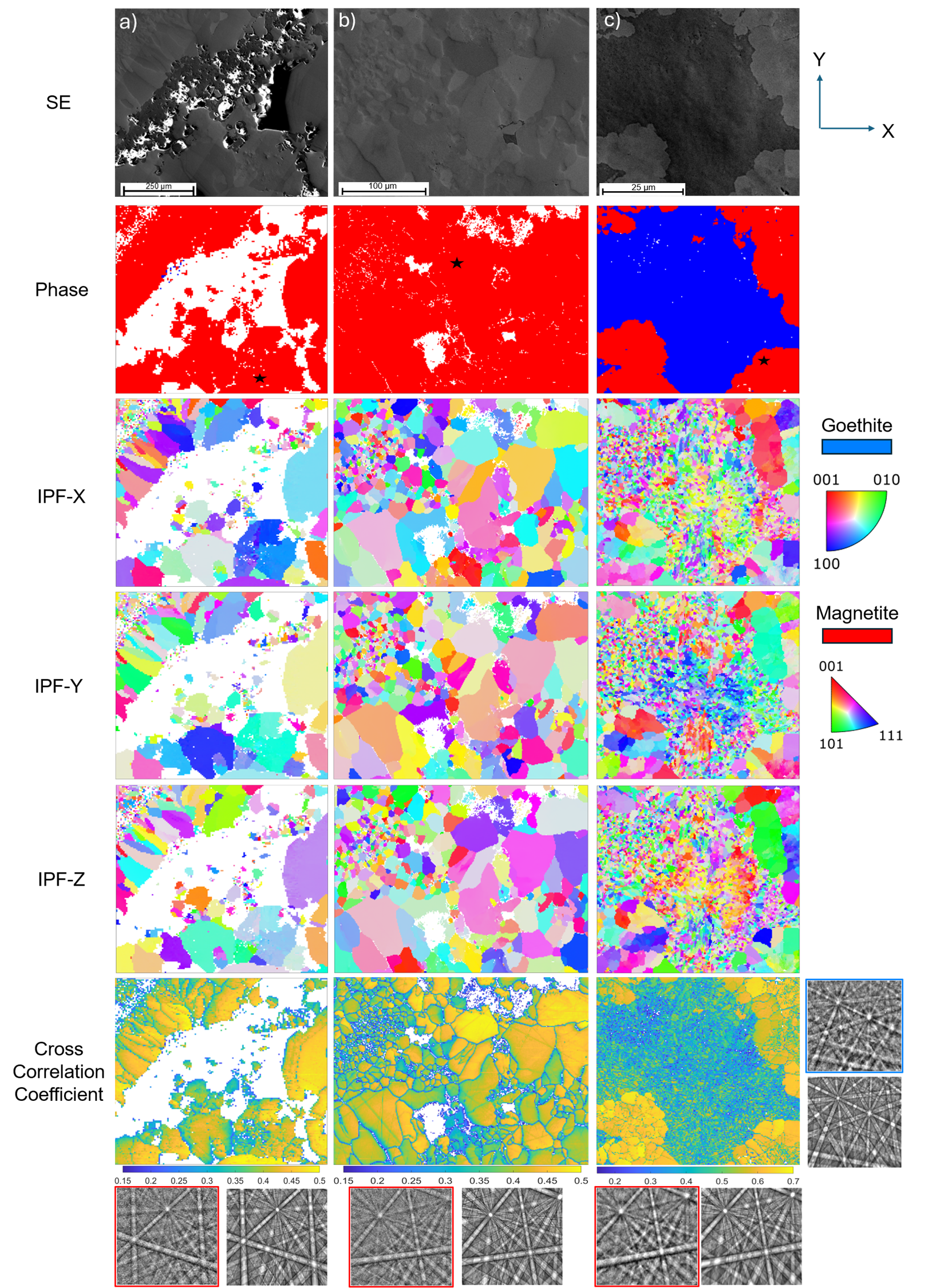}
    \caption{SE and EBSD data captured for phase identification at three regions of interest. EBSD data includes phase colour map, Inverse Pole Figure (IPF) for x, y, and z direction, cross correlation coefficient, and the electron backscatter pattern (EBSP) for magnetite (red boxes) and goethite (blue box) compared to their dynamically simulated EBSP (no box).}
    \label{Fig:PhaseID}
\end{figure}

 Simultaneous EBSD and EDS was performed at three regions on the sample surface with distinct and different phases. The region shown in Fig. \ref{Fig:PhaseID} a) contained large elongated grains, a region of BSE-darker phase, and voids with crystalline phases inside. The matrix in this region was comprised of magnetite and no distinct phase identification was possible for the crystalline phases or darker region. The inverse pole figures (IPF) from EBSD also showed no local microtexture in the microstructure. However, there is a significant change in size and morphology of the magnetite grains as one crosses the BSE-darker region. Comparison of the experimentally obtained electron backscatter pattern (EBSP) and a dynamically simulated EBSP for magnetite also showed significant confidence in the magnetite identification for the matrix. EBSD of the second region (Fig. \ref{Fig:PhaseID} b) showed a similar result to the first, with the matrix being identified as magnetite and low confidence for identification of other phases. However, a significant difference in the matrix grain structure was seen in this region. The magnetite grains in this region were seen to vary in size with smaller grains being closer to the high Ni content region shown in Fig. \ref{Fig:LargeMapChemistry}. Analysis of the final region showed both magnetite and goethite present on the sample surface. Though the grain structuring of goethite was not well resolved due to the concave topography in the region.

\subsection{Mixed Phase Inclusion}

\par The brecciated mixed phase inclusion shown in Fig. \ref{Fig:LargeStitchedMap} b) was also investigated using XPS, EBSD, and EDS. Through XPS and EDS, it was found that the inclusion had a similar chemical composition to the matrix, and it is predominantly an Fe and Ni oxide with peaks corresponding to NiO and Fe\textsubscript{2}O\textsubscript{3}, as described above.

\par
\begin{figure}[!hbt]
    \centering
    \includegraphics[width=0.95\linewidth]{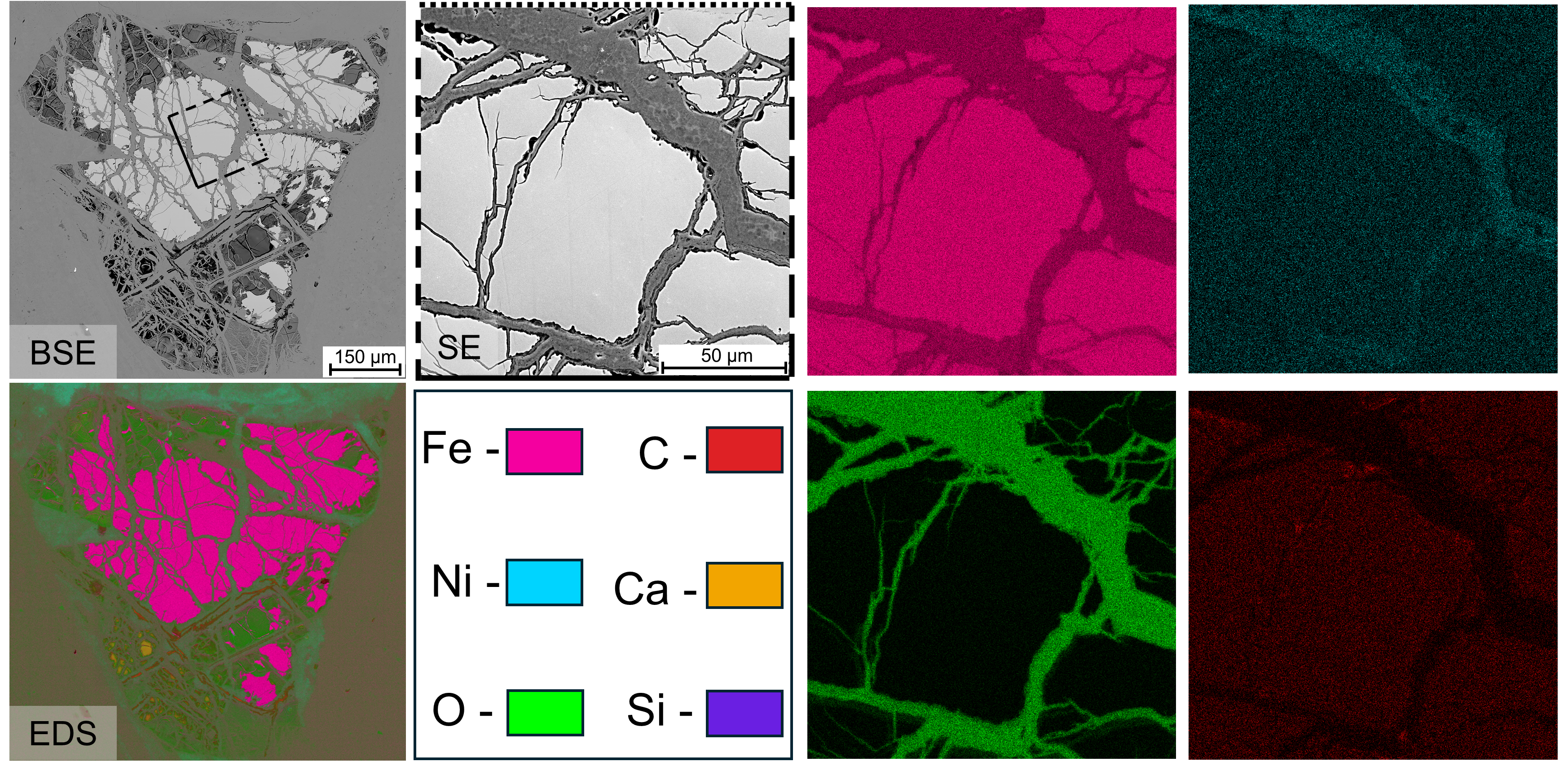}
    \caption{BSE, EDS, and SE images of large inclusion in the Nantan fragment. Overview of the inclusion is given on the left with a stitched EDS map. An increased magnification SE and EDS images are provided to the right of the area shown with the black box.}
    \label{Fig:InclusionEDS}
\end{figure}

 EDS analysis of this region showed relative enrichment of C and Fe in the brecciated phase which appears lighter under BSE (Fig. \ref{Fig:InclusionEDS}). The vein-like structure and BSE-darker phase surrounding the brecciated phase were found to contain Fe, Ni, and O. Ca was also found to be present in small pockets in the inclusion, similar to the previously shown large area EDS mapping. Based on these results, the BSE-lighter phase may be composed of an Fe-Ni carbide (cohenite) while the veins-like structure are extensions of the sample matrix and are composed of an iron oxide and nickel oxide mixture. There was also a notable drop in O concentration and increase in the C concentration within the inclusion compared to the matrix which is attributed to the large cohenite grains present. 

\par
\begin{figure}[!hbt]
    \centering
    \includegraphics[width=\linewidth]{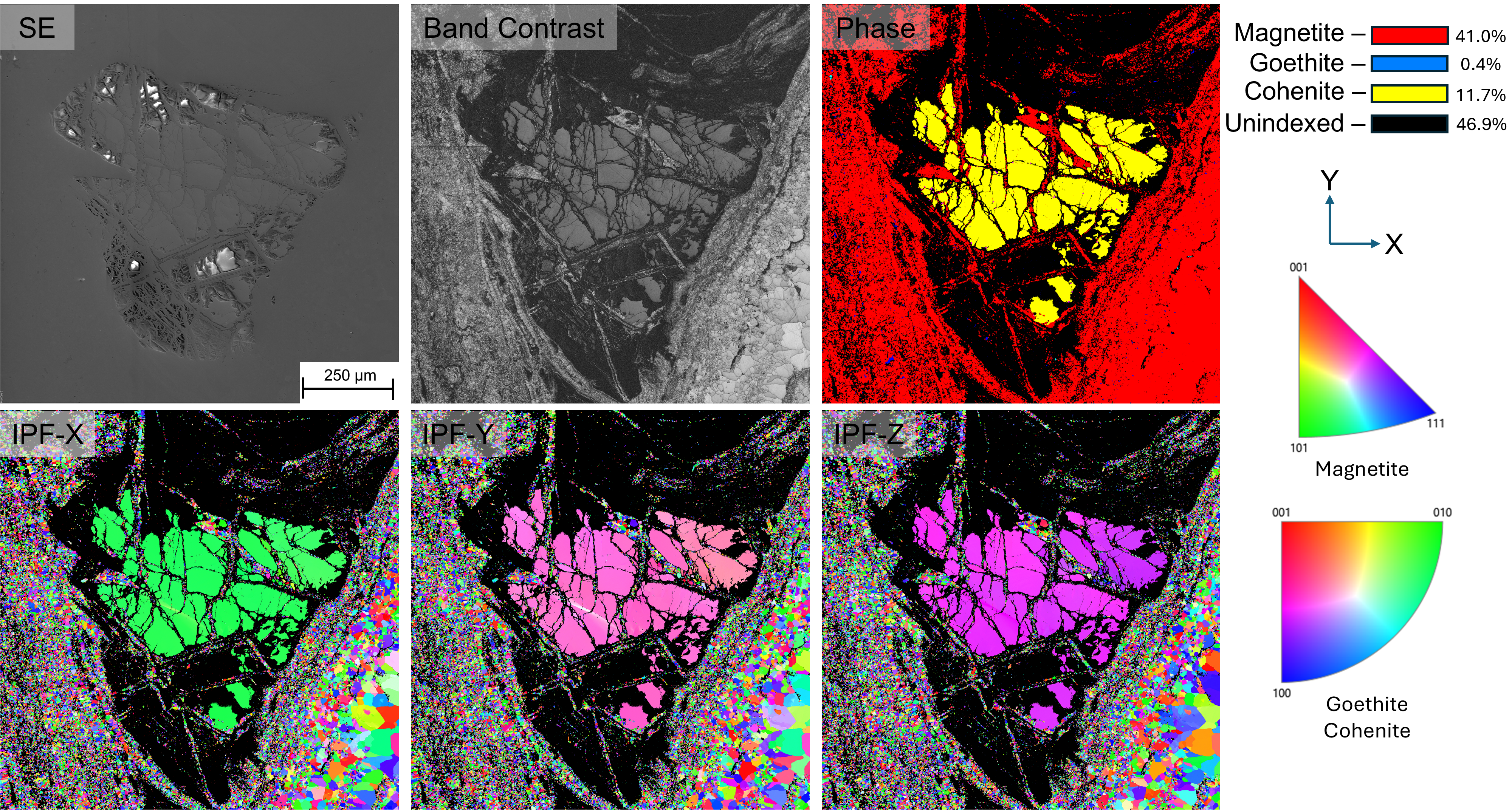}
    \caption{SE, Band Contrast, phase colour, and IPF maps of the large inclusion within the Nantan fragment captured using EBSD. The phase colour map shows the large fracture phase in the centre to be cohenite. The IPF maps show a rapid transition of grain size and that all the cohenite grains are in the same orientation.}
    \label{Fig:InclusionEBSD}
\end{figure}

 EBSD analysis of this region showed a unique grain structure where the phase with the highest backscatter coefficient (white in the BSE image) for the inclusion was a single grain of near-uniform orientation. This grain is brecciated, with an `infill' of fine grain vein-like structures extending outward from the matrix (Fig. \ref{Fig:InclusionEBSD}). The surrounding matrix is made up of large randomly oriented grains that decrease in size nearest the inclusion. The vein-like structure can also be seen within the grain structure of the matrix with the grains of the vein-like structure being relatively uniformly structured along the path of the vein while the surrounding grains of the matrix are equiaxed but random. Additionally, the phase colour map showed that the matrix composition remained magnetite near the inclusion, while the white phase were composed of cohenite. The other phases within the inclusion were not able to be identified using EBSD.

\par The elemental composition of the matrix and inclusion determined  using XPS, EDS, and XRF, are given in Table \ref{Tbl:CompositionResults}. 

\begin{table}[!hbt]
        \centering
        \begin{tabular}{|c|c|c|c|}
        \hline
        XPS (at\%) & Matrix (High Ni) & Matrix (Low Ni) & Inclusion \\ \hline
        Fe & 29.29$\pm$0.29 & N/A & 32.43$\pm$0.59 \\ \hline
        Ni & 2.67$\pm$0.13 & N/A & 0.61$\pm$0.16 \\ \hline
        O & 66.31$\pm$0.30 & N/A & 54.99$\pm$0.52 \\ \hline
        C & 1.43$\pm$0.13 & N/A & 11.70$\pm$0.26 \\ \hline
        Cl & 0 & N/A & 0 \\ \hline
        Si & 0 & N/A  & 0 \\ \hline
        Co & 0 & N/A & 0 \\ \hline
        EDS (at\%) & Matrix (High Ni) & Matrix (Low Ni) & Inclusion \\ \hline
        Fe & 27.8$\pm$0.1 & 31.7$\pm$0.1 & 37.7$\pm$0.1 \\ \hline
        Ni & 2.8$\pm$0.1 & 0.2$\pm$0.1 & 0.8$\pm$0.1 \\ \hline
        O & 55.6$\pm$0.3 & 54.8$\pm$0.3 & 33.4$\pm$0.1 \\ \hline
        C & 13.4$\pm$0.1 & 13.1$\pm$0.1 & 27.8$\pm$0.1 \\ \hline
        Cl & 0 & 0 & 0 \\ \hline
        Si & 0.3$\pm$0.1 & 0.2$\pm$0.1 & 0.2$\pm$0.1 \\ \hline
        Co & 0 & 0 & 0 \\ \hline
        XRF (at\%) & Matrix (High Ni) & Matrix (Low Ni) & Inclusion \\ \hline
        Fe & 87.73$\pm$0.22 & 98.29$\pm$0.24 & 96.93$\pm$0.24 \\ \hline
        Ni & 10.08$\pm$0.12 & 0.17$\pm$0.02 & 1.78$\pm$0.05 \\ \hline
        O & 0 & 0 & 0 \\ \hline
        C & 0 & 0 & 0 \\ \hline
        Cl & 0.07$\pm$0.06 & 0.02$\pm$0.05 & 0.34$\pm$0.06 \\ \hline
        Si & 0.27$\pm$0.14 & 0.21$\pm$0.13 & 0.12$\pm$0.13 \\ \hline
        Co & 0.94$\pm$0.05 & 0.42$\pm$0.05 & 0.51$\pm$0.05 \\ \hline
        Phases & \begin{tabular}[c]{@{}c@{}}Magnetite (I), \\ NiO (H), Maghemite (I)\end{tabular} & \begin{tabular}[c]{@{}c@{}}Magnetite (I) \\ Goethite (O), Maghemite (I)\end{tabular} & \begin{tabular}[c]{@{}c@{}} Maghemite (I), \\ Cohenite (O), NiO (H)\end{tabular} \\ \hline
        \end{tabular}
    \caption{Elemental composition of the Nantan sample found using XPS, EDS, and XRF at the three regions of interest identified using XRF in Fig. \ref{Fig:LargeMapChemistry} c). XPS was only captured at the high Ni section of the matrix and the inclusion. Crystal structures: (I) = Inverse spinel, (O) = Orthorhombic, (H) = Halite.}
    \label{Tbl:CompositionResults}
\end{table}

\section{Discussion}

\subsection{Phases and Composition of the Matrix}

\par Correlative microstructural analysis of the Nantan meteorite fragment using EDS, XRF, and XPS confirmed the elemental composition to be a majority Fe, Ni, and O, with no metallic Fe/Ni alloy found. The distribution of Ni was found using EDS, XRF, and XPS to be heterogeneous with large areas of higher Ni concentration ($\geq$\SI{2.6}{at\percent Ni}), as shown in Fig. \ref{Fig:LargeMapChemistry}. Other elements common in weathering products (Si, Ca and C) were also found in this meteorite fragment. It is believed this Nantan fragment underwent significant weathering since its fall due to the lack of non-oxide, e.g. kamacite or taenite, grains. The combination of data from XRD, EDS, and XPS suggests that the Nantan fragment was primarily composed of magnetite, with maghemite, goethite, and NiO intermixed within the matrix.

\par The microstructure of the matrix was seen to form two sets of distinct regions. The first identified by size of the grains and the grain structure (Fig. \ref{Fig:LargeStitchedMap} a), the second being band-like regions that is seen from the elemental composition (Fig. \ref{Fig:LargeMapChemistry} a). The grain size regions generally separate the sample down the middle, with a fine equiaxed grain structure on the left side and an elongated grain structure with larger grains on the right with crystallized phases present (Fig. \ref{Fig:LargeStitchedMap} c) and dark precipitates (Fig. \ref{Fig:LargeStitchedMap} d). It is believed that the dark precipitates and crystallized phases are goethite, as there is a correlation of goethite within the XRD results of the sample and the experimental EBSPs obtained at these regions match the simulated patterns. 

\par Additionally, XRD analysis of the polished sample surface showed strong correlation with magnetite, maghemite and goethite, and a weak correlation with lepidocrocite. Each of which are common products of aqueous weathering in meteorites. The weaker peak intensities of goethite in the experimentally obtained diffractogram is likely due to limited amounts of goethite being exposed at the surface compared to magnetite. The presence of lepidocrocite cannot be confidently confirmed from the experimental XRD data given the weak peak intensities or the EBSD data as the EBSPs did not provide a significant confidence for lepidocrocite over other iron oxyhydroxide weathering phases. 

\subsection{Ni Concentration Separated Regions}

\par The band-like elemental composition regions were seen to be separated by the Ni concentration, with the high Ni regions having concentrations ($\geq$\SI{2.6}{at\percent}) and the low regions ($\leq$\SI{0.9}{at\percent}) Ni. These regions are believed to have been formed from a former Widmanstätten pattern present within the meteorite fragment prior to aqueous weathering, with the low and high Ni regions being former kamacite and plessite (kamacite+taenite), respectively. 

\par Quantitative analysis of the high Ni region (green circle, Fig. \ref{Fig:LargeMapChemistry} c) showed Fe:Ni ratios of 10:1 (27.8:2.8), 11:1 (29.29:2.67), and 9:1 (87.73:10.08), for EDS, XPS, and XRF respectively, or an average of 10:1. Previous work by \emph{Lee and Bland} has found the Fe:Ni ratios of weathered products from kamacite and taenite to be $\sim 9-18$ and $\sim 20-25$, respectively \cite{Lee2004}. The Fe:Ni ration found in this work matches well with previously reported values for weathering product of kamacite \cite{Binzel2006,Lee2004,Palmer2011}. The remaining higher Ni areas are then suggested to be plessite regions with higher amounts of kamacite, given the similarity of the weathered surface's Fe:Ni ration to that of weathered kamacite. Sadly, the specific morphology of the former plessite cannot be determined from the current information and would require additional investigation.

\par The weathering on the phases in this region is believed to differ for the former plessite and taenite regions, with the plessite experiencing rapid dissolution of the kamacite into magnetite and the taenite grains undergoing the mechanism given in Eqn. \ref{Eqn:AkaganeiteDecomA} and \ref{Eqn:AkaganeiteDecomB}. This mechanism results in the formation of maghemite and NiO within the high Ni region, alongside the magnetite generated from the kamacite. The low Ni region would have undergone similar weathering mechanisms, though with the generation of significantly greater goethite. From the results of XRF and EBSD, there appears to be a correlation between the Cl concentration with the presence of goethite in this Nantan fragment. The higher amount of goethite in this region may be due to an increased presence of akaganéite that was catalyzed by the Cl\textsubscript{-} present in the groundwater and which decomposed over time.

\subsection{Grain Structure Separated Regions}

\par There is a clear border defined by the grain structure of the sample which generally travels down the centre with a varying width of 100-\SI{200}{$\mu m$}, with smaller (approx. \SI{5}{$\mu$m}) equiaxed grains on the left and large  ($\geq 10 \mu m$) elongated grains on the right. The grain structure boundary does not appear to be related to the composition of the major elements (Fe, Ni, O) within the sample. Rather, the matrix grain sizes appear larger in areas surrounding goethite grains and where localized areas of elevated concentrations of Cl are present (Fig. \ref{Fig:LargeMapChemistry} c). 

\par Areas or "hot spots" of concentrated Cl are not unusual within weathered meteorites \cite{Buchwald1989, VanGinneken2022}. These are typically remnants of the Cl which entered the meteorite from groundwater sources during the weathering process. This Cl is then transported out of the meteorite or retained as akaganéite or hibbingite. It may be that the meteorite was originally exposed to groundwater on the right side of the sample. The Cl present in the groundwater then attacked the kamacite grains present, forming akaganéite and/or hibbingite throughout the outer volume of the fragment. Some fraction of these phases then decomposed and formed goethite, maghemite, and magnetite. The rest remaining as meta-stable reaction products and hot spots of Cl concentration. A similar post-weathering Cl distribution was also seen by Ginneken \emph{et al.} when artificially weathering a chondrite meteorite, where the Cl-containing species formed sub-spherical and tubular shells \cite{VanGinneken2022}. It is possible that similarly structured phases exists within the Nantan fragment investigated here, however further work would need to be performed to investigate the 3D microstructure of the sample.

\par The discovered distinct grain structure regions are then a product of the weathering mechanism which occurred. Grains on the right side of the boundary interacted with the Cl and formed akaganéite, which then gradually decomposed into goethite, maghemite, and magnite. This gradual decomposition would allow for the formation of larger grains as former grains transformed. The kamacite grains on the left side of the boundary likely underwent rapid weathering and decomposition to magnetite, forming many but small grains as they simultaneously formed.

\par No akaganéite was identified on the investigated sample surface using XPS, EDS, and EBSD. However, this does not mean the Cl detected by XRF should be disregarded. It is possible that akaganéite exists deeper within the Nantan fragment, with high Cl quantities within or surrounding the akaganéite. As seen by Buckwald and Clarke \cite{Buchwald1989}, the Cl content of akaganéite is higher in the deep interior of meteorites when it is a crack filling, with cracks extending up to 15 mm. Identification of sub-surface species is effectively impossible for EBSD and XPS given their penetration depths, (\SI{150}{nm} in low density material) and ($\sim1-$\SI{5}{nm} in iron) respectively \cite{Wisniewski2017, Greczynski2022}, unlike XRF which is capable of reaching millimetre depths in ideal conditions \cite{Boone2010}.



\subsection{Inclusion and Vein-like Microstructure Formation}

\par The elemental composition and microstructure of the inclusion were found to be unique compared to the rest of the sample. In Fig. \ref{Fig:InclusionEDS}, and from the correlative analysis, there are four distinct regions and phases; 

\begin{itemize}
    \item Brecciated white inclusions - Fe, C, Ni ((Fe,Ni)\textsubscript{3}C)
    \item Vein-like structure - Fe, Ni, O (Fe\textsubscript{3}O\textsubscript{4}, $\gamma$-Fe\textsubscript{2}O\textsubscript{3}, NiO)
    \item Light grey cracked inclusions in voids - Ca, C, O (CaCO\textsubscript{3})
    \item Dark grey inclusions - Fe, C, O (FeCO\textsubscript{3})
\end{itemize}

\par It is likely that each of the carbide inclusion grains present in this region originated from a singular parent grain, as the IPF maps in Fig. \ref{Fig:InclusionEBSD} show each grain shares a common crystal orientation. The elemental composition and EBSD indexing indicate that this carbide phase is cohenite. The vein-like phase then formed as an extension of the high Ni section of the matrix through aqueous weathering and infiltration of the solution into the cracked cohenite, which is a common occurrence seen during oxidation of meteorites \cite{Binzel2006, Gooding1986}. This resulted in the formation of magnetite, maghemite, and NiO between the cohenite grains, via the mechanisms discussed previously. The Fe and Ca-carbonates are secondary minerals which likely formed as part of the aqueous weathering process and incorporation of CO\textsubscript{3}\textsuperscript{2-} from the local chemical environment \cite{Rubin1997}.

\par The grain structure seen in Fig. \ref{Fig:InclusionEBSD} can be attributed to the simultaneous grain formation during weathering and the grain refinement of Fe with the increase of Ni content. Weathering of the original kamacite matrix would have occurred rapidly over large areas of the material surface, causing a large number of grains to form simultaneously. This in combination with the known effect of Ni to reduce the grain size of iron alloys \cite{Pratomo2019, Tong2021} results in the microstructure observed. It is not known if the increased presence of Ni itself or the weathering mechanism is the dominant contribution to this reduced grain size. However, a clear difference was seen in the grain structure of the high and low Ni sections of the matrix. 

\par Lastly, it is suggested that the inclusion region is a cross-section of a much larger feature within the meteorite fragment. The vein-like structure was seen to extend below the inclusion region using EDS (Fig. \ref{Fig:LargeMapChemistry} a), with the Ni distribution having a similar fractured appearance to the inclusion. Additionally, the species present within pores of the inclusion were seen using SEM to extend further into the sample surface with varying depths.

\section{Conclusion}

\par The Nantan fragment was found to have experienced extreme weathering, resulting in the sample to be composed of primarily magnetite, with NiO, goethite, and maghemite present. The matrix was seen to form two sets of distinct regions, a grain structure separated regions and a Ni concentration separated regions. The Ni concentration region was seen using EDS, XPS, and XRF with a high ($\geq$\SI{2.6}{$at\%$}) and a low ($\leq$\SI{0.9}{$at\%$}) Ni concentration region. EDS determined the elemental composition in the high and low Ni regions of the matrix to be \SI[separate-uncertainty = true]{27.8(0.1)}{Fe}-\SI[separate-uncertainty = true]{2.8(0.1)}{Ni}-\SI[separate-uncertainty = true]{0.3(0.1)}{Si}-\SI[separate-uncertainty = true]{55.6(0.3)}{O}-\SI[separate-uncertainty = true]{13.4(0.1)}{C} ($at\%$) and \SI[separate-uncertainty = true]{31.7(0.1)}{Fe}-\SI[separate-uncertainty = true]{0.2(0.1)}{Ni}-\SI[separate-uncertainty = true]{0.2(0.1)}{Si}-\SI[separate-uncertainty = true]{54.8(0.3)}{O}-\SI[separate-uncertainty = true]{13.1(0.1)}{C} ($at\%$), respectively. 

\par These high and low Ni concentration regions are believed to have formed from former plessite and kamacite present in the sample, respectively. The low Ni regions are believed to have been formed from the aqueous alternation of kamacite into the chlorine stabilized akaganéite, which then decomposed into magnetite, maghemite and goethite. The high Ni region is believed to have undergone two simultaneous weathering mechanisms, rapid dissolution and precipitation of the kamacite into magnetite and formation of maghemite and NiO through aqueous alteration of taenite.

\par The grain structure separated regions were found to separated by a visible surface boundary of 100-\SI{200}{$\mu m$} width and are believed to correlate with the presence of Cl. On one side of this boundary, smaller grains (approx. \SI{5}{$\mu m$}) were found with low but generally dispersed Cl content. The other side had larger grains ($\geq$ \SI{10}{$\mu m$}) and concentrated regions of Cl rather than dispersed. It was also found that significant amounts of goethite was present around areas of elevated Cl concentrations. It is suggested that these spots of elevated Cl concentration are from subsurface akaganéite or hibbingite left over from the weathering mechanism, with akaganéite near the analysis surface decomposing into goethite, magnetite, and maghemite. Goethite was not found within the smaller grain region and it is suggested that this region formed through rapid weathering and decomposition to magnetite. Other signs of weathering were also found in a large inclusion present at the analysis surface. 

\par This inclusion was primarily composed of cohenite that appeared to be cracked, likely due to an impact the meteorite experienced. The cracks between the cohenite grains were found to be filled with a vein-like structure extending from the matrix and with small phases of Fe and Ca carbonate. The vein-like structure is believed to have formed during the aqueous alteration that formed the surrounding microstructure, with the aqueous solution penetrating into the cracks of the cohenite and depositing NiO, maghemite, and magnetite. 

\par Finally, it is suggested that this large inclusion is a small cross-section of a much larger feature within the meteorite body. The distribution of Ni was shown to form a silhouette that extended below the inclusion into a tip and the pores within the inclusion appear to extend further below the analyzed surface.

\section*{Acknowledgements}

We acknowledge the following funding support: Natural Sciences and Engineering Research Council of Canada (NSERC) through their Discovery Grant program [Discovery grant: RGPIN-2022–04762, ‘Advances in Data Driven Quantitative Materials Characterisation’], the Vanier Canada Graduate Scholarship, the Canadian Research Chairs program, B.C. Knowledge Development Fund (BCKDF) and Canadian Foundation for Innovation through the Innovation Fund (CFI-IF) (\#39798: AM+) and the John R. Evans Leaders Fund (CFI-JELF) [\#43737, 3D-MARVIN]. We thank the Fipke Laboratory for Trace Element Research (FiLTER) at UBC Okanagan for provision of equipment for the XRF analysis. From Thermo Scientific, we acknowledge Helen Oppong-Mensah and Chris Stephens for the assistance with the MAPS software. Electron microscopy was performed at the Electron Microscopy Laboratory (Materials Engineering, UBC).

\section*{CRediT}

\textbf{Graeme J. Francolini:} Conceptualization, Formal Analysis, Investigation, Methodology, Visualization, Writing - Original draft, Writing - review \& editing. \textbf{Paul Mack:} Investigation, Resources. \textbf{Brendan V. Dyck:} Investigation, Resources,  Writing - review \& editing. \textbf{Ben Britton:} Formal Analysis, Funding acquisition, Resources, Supervision, Writing - review \& editing.

\section*{Data Availability}
Data will be released to Zenodo when the article is accepted for publication, and a DOI will be inserted here.
\bibliographystyle{elsarticle-num}

\bibliography{article.bib}

\end{document}